\documentclass[11pt]{article}
\usepackage{amssymb}
\begin{document}
\setlength{\oddsidemargin}{0cm} \setlength{\evensidemargin}{0cm}
\baselineskip=20pt

\begin{center} {\Large\bf From Rota-Baxter Algebras to Pre-Lie Algebras}
\end{center}

\bigskip

\begin{center}  { \large Huihui An$^{a}$ \footnote{ E-mail address:
anhh@mail.nankai.edu.cn}\qquad \large Chengming ${\rm Bai}^{b}$}
\end{center}

\begin{center}{\it a. Department of Mathematics \& LPMC, Nankai
University, Tianjin 300071, P.R. China }
\end{center}

\begin{center}{\it b. Chern Institute of Mathematics \& LPMC, Nankai University, Tianjin
300071, P.R. China}\end{center}

\bigskip

\centerline{\large\bf   Abstract } \vspace{0.2cm}

Rota-Baxter algebras were introduced to solve some analytic and
combinatorial problems and have appeared in many fields in
mathematics and mathematical physics. Rota-Baxter algebras provide a
construction of pre-Lie algebras from associative algebras. In this
paper, we give all Rota-Baxter operators of weight 1 on complex
associative algebras in dimension $\leq 3$ and their corresponding
pre-Lie algebras.

\vspace{0.4cm}


\vspace{0.4cm}


\newpage

\section {Introduction}\setcounter{equation}{0}

A Rota-Baxter algebra is an associative algebra $A$ over a field
${\bf F}$ with a linear operator $R:A\rightarrow A$ satisfying the
Rota-Baxter relation:
$$R(x)R(y)+ \lambda R(xy)=R(R(x)y+xR(y)),
\forall x,y \in A.\eqno (1.1)$$ Here $\lambda\in{\bf F}$ is a fixed
element which is called the weight. Obviously that for any $\lambda
\neq 0$, $R \rightarrow {\lambda}^{-1}R$ can reduce the Rota-Baxter
operator $R$ of weight $\lambda$ to be of weight $\lambda =1$.

Rota-Baxter relation (1.1) first occurred in the  work of G. Baxter
in 1960  to solve an analytic problem ([Bax]), based on a  paper
written by F. Spitzer ([Sp]) in 1956. In fact, the Rota-Baxter
relation (1.1) generalizes the integration-by-parts formula.  G.-C.
Rota ([R1-R4]), F.V. Atkinson ([At]) and P. Cartier ([Ca])
contributed important results. In particular, it was G.-C. Rota who
realized its importance in combinatorics and other fields in
mathematics ([R1-R2]). Since then, it has been related to many
topics in mathematics and mathematical physics. For example,
Rota-Baxter algebras appeared in connection with the work of A.
Connes and D. Kreimer on renormalization theory in perturbative
quantum field theory ([CK2-3]), see [FG] for more details. It is
also related to J.-L. Loday's dendriform algebras ([Lo], [LR]), as
well as to M. Aguiar's associative analogue of the classical
Yang-Baxter equation ([Ag1-3]).

However, it is difficult to construct examples of Rota-Baxter
algebras. Basically there are two ways to construct Rota-Baxter
algebras. One way is to use the free Rota-Baxter algebras which in
some sense  are the ``biggest" examples. There are a lot of
references on the study of free Rota-Baxter algebras ([Ca], [R1],
[EG2], [GK1-2] and the references therein). The other way is to get
 concrete examples in low dimensions, which is the main content of
this paper. Although there has already existed certain works on
(finite-dimensional) Rota-Baxter algebras, e.g. [Deb], [Der],
[Mi1-2], [N], to our knowledge, there has  been no
``classification'' in low dimensions yet. We will give all
Rota-Baxter algebras in dimension $\leq 3$. Though our study depends
on  direct computation through example one by one, these examples
will be regarded as a guide for  further development.

An application of Rota-Baxter (associative) algebras is to get some
new algebraic structures. We mainly mention two classes of algebraic
structures related to Rota-Baxter algebras in this paper. One class
of algebras are the Loday's dendriform algebras ([Lo], [LR]).
Dendriform algebras are equipped with an associative product which
can be written as a linear combination of nonassociative
compositions. These notions are motivated by the natural link
between associative algebras and Lie algebras. By the work of M.
Aguiar, P. Leroux and K. Ebrahimi-Fard ([Ag1], [E1-2], [Le1-2]) the
close relation of these new types of algebras to Rota-Baxter
algebras as well as Nijenhuis algebras and differential algebras was
established.

The other class of algebras are the pre-Lie algebras (or have other
names such as left-symmetric algebras, Vinberg algebras and so on).
Pre-Lie algebras are a class of nonassociative algebras coming from
the study of convex homogeneous cones, affine manifolds and
deformations of associative algebras ([Au], [G], [Ki], [Me], [V]).
As it was pointed out in [CL], the pre-Lie algebra ``deserves more
attention than it has been given". It has also appeared in many
fields in mathematics and mathematical physics, such as complex and
symplectic structures on Lie groups and Lie algebras ([AS], [Ch],
[Sh]), integrable systems ([SS]), classical and quantum Yang-Baxter
equations ([Bo], [ES], [GS], [Ku1-2]), Poisson brackets and
infinite-dimensional Lie algebras ([BN], [GD], [Z]), vertex algebras
([BK]), quantum field theory ([CK1]) and operads ([CL]). In
particular, an important role has been played by pre-Lie algebras in
mathematical physics, especially the work of Connes-Kreimer on
pre-Lie algebra structure on Feynman diagrams by the
insertion-elimination operations (see [CK4] for a detailed
interpretation). The same can be said of Rota-Baxter algebras. The
connection of these two roles is still not clear, which might be
clarified by careful study on the relation between Rota-Baxter
algebras and pre-Lie algebras, as we try to do in this paper.

Since there is no suitable (matrix) representation theory of pre-Lie
algebras due to their nonassociativity, it is natural to consider
how to construct them from some  algebraic structures which we have
known. This is the ``realization theory". We have already obtained
some experience. For example, a commutative associative algebra
$(A,\cdot)$ and its derivation $D$ can define a Novikov algebra
$(A,*)$ (which is a pre-Lie algebra with commutative right
multiplication operators) by ([GD], [BM1-2]):
$$x*y=x\cdot Dy,\;\;\forall x,y\in A.\eqno (1.2)$$
An analogue of the above construction in the version of Lie algebras
is related to the classical Yang-Baxter equation. In fact, a Lie
algebra $({\cal G},[\;,\;])$ and a linear map $R:{\cal G}\rightarrow
{\cal G}$ satisfying
$$[R(x),R(y)]=R([R(x),y]+[x,R(y)]), \;\;\forall x,\;y\in {\cal
G}\eqno (1.3)$$ can define a pre-Lie algebra $({\cal G},
*)$ by ([BM3], [GS], [Ku3], [Me])
$$x*y=[R(x),y],\;\;\forall x,\;y\in {\cal G}.\eqno (1.4)$$
Equation (1.3) is just the operator form of classical Yang-Baxter
equation on a Lie algebra which was given by  M.A.
Semenov-Tyan-Shanskii in [Se]. Obviously it also can be regarded as
a Rota-Baxter operator of weight zero on the Lie algebra ${\cal G}$.
In fact, as it was mentioned in [EGK] and [EG2], the Rota-Baxter
relation (1.1) on associative algebras can be naturally extended to
be on Lie algebras.

It is natural to consider the construction of pre-Lie algebras from
(noncommutative) associative algebras. The answer is the
construction from Rota-Baxter algebras.  Let $(A,\cdot)$ be an
associative algebra and $R$ be a Rota-Baxter operator. If the weight
$\lambda=0$, then from equations (1.3) and (1.4), it is obvious that
the product
$$x*y=R(x)\cdot y-y\cdot R(x),\;\;\forall x,y\in A \eqno (1.5)$$
defines a pre-Lie algebra. When the weight $\lambda=1$, we can  see
that the product
$$x*y=R(x)\cdot y-y\cdot R(x)-x\cdot y,\;\;\forall x,y\in A \eqno (1.6)$$
defines a pre-Lie algebra (see Corollary 2.7). In fact, there are
two approaches to both equations (1.5) and (1.6). One approach is
from the relation between pre-Lie algebras and the operator form of
the (modified) classical Yang-Baxter equation given by I.Z.
Golubchik and V.V. Sokolov in [GS]. The other approach is from the
relation between dendriform dialgebras and Rota-Baxter algebras and
pre-Lie algebras given by M. Aguiar and K. Ebrahimi-Fard ([Ag1],
[E1-E2]). It is also natural to consider which kind of pre-Lie
algebras can be obtained from Rota-Baxter algebras.

Note that for a commutative associative algebra, the inverse of an
invertible derivation is just a Rota-Baxter operator of weight zero.
So we would like to point out that in the above three algebraic
constructions (commutative associative algebras, Lie algebras and
associative algebras) of pre-Lie algebras, the corresponding linear
transformations (derivations, operators satisfying classical
Yang-Baxter equation and Rota-Baxter operators) have more or less
relations to Rota-Baxter operators.

We have given a detailed study of Rota-Baxter operators on pre-Lie
algebras of weight zero in [LHB]. A more remarkable property is that
for any such Rota-Baxter pre-Lie algebra, equation (1.5)  can also
define a pre-Lie algebra which is called the double of the former
([LHB]). Therefore, any pre-Lie algebra with its Rota-Baxter
operator (of weight zero) and its doubles  can construct a close
category. We would like to point out that there is another different
double construction of Rota-Baxter algebras defined by Ebrahimi-Fard
in [EGK], that is, for any Rota-Baxter algebra $(A,R)$, there is a
new Rota-Baxter algebra $(A_R,R)$ which is called the double of
$(A,R)$ in [EGK], where the product in $A_R$ is given by
$$x*_Ry=R(x)y+xR(y)-xy,\;\;\forall\; a, b\in A.\eqno (1.7)$$
Moreover, all Rota-Baxter operators of weight zero on associative
algebras in dimension $\leq 3$ were given in [LHB] too.

In this paper, we study the Rota-Baxter operators of weight $\lambda
=1$ on associative algebras. It is easy to see that this Rota-Baxter
operator is still a Rota-Baxter operator on the induced pre-Lie
algebra given by equation (1.6) ([EGP]). The paper is organized as
follows. In section 2, we give some fundamental results and examples
on Rota-Baxter algebras and pre-Lie algebras. In section 3, we give
all Rota-Baxter algebras on 2-dimensional complex pre-Lie algebras,
and in the associative cases, we give their corresponding pre-Lie
algebras. In section 4, we give all Rota-Baxter algebras on
3-dimensional complex associative algebras and their corresponding
pre-Lie algebras. In section 5, we give some  discussion and
conclusions.

Throughout this paper, the Rota-Baxter operator is of weight
$\lambda=1$ and all algebras are of finite dimension and over the
complex field ${\bf C}$, unless otherwise stated. $\langle\ |\
\rangle$ stands for an associative  algebra with a basis and nonzero
products at each side of  `` $|$ ".

\section{Preliminaries and some examples}\setcounter{equation}{0}

  Let $A$ be an associative algebra. For any $x,y\in A$, the
commutator $[x,y]=xy-yx$ defines a Lie algebra. We denote the set of
all Rota-Baxter operators on $A$ of weight $\lambda=1$ by ${\rm
RB}(A)$. Then the following conclusion is obvious (cf. [E1], [EGP],
[EG1], etc.).

{\bf Lemma 2.1}\quad  Let $(A,\cdot)$ be an associative algebra.

(1) A linear operator $R\in {\rm RB}(A)$ if and only if $1-R\in {\rm
RB}(A)$, where $1$ is the identity map. In particular, $0,1\in {\rm
RB}(A)$.

(2) Let $(A,*)$ be an algebra given by
$$x*y=R(x)\cdot y+x\cdot R(y)-x\cdot y,\;\;\forall\; x,y\in A.\eqno
(2.1)$$ Then $(A,*)$ is an associative algebra  and $R$ is still a
Rota-Baxter operator of weight 1 on $(A,*)$.

(3) If $R\in {\rm RB}({A})$, then $B=1-2R$ satisfies
$$[B(x), B(y)]+[x,y]=B([B(x),y]+[x, B(y)]),\;\; \forall x, y \in A.\eqno (2.2)$$

(4) Let $A'$ denote the algebra defined by a product
$(x,y)\rightarrow x\circ y$ on $A$ which satisfies $x\circ y=y\cdot
x$ for any $x,y\in A$, then $A'$ is still an associative algebra and
${\rm RB}(A)={\rm RB}(A')$.

(5) If $R \in {\rm RB}({A})$ and $R^2=R$, then for any $\alpha\in
{\bf F}$, $N_{\alpha}=(1+\alpha)R-\alpha$ satisfies the following
Nijenhuis relation ([CGM], [Le1-2])
$$N_{\alpha}(x)N_{\alpha}(y)+ N_{\alpha}^2(xy)=N_{\alpha}(N_{\alpha}(x)y+xN_{\alpha}(y)), \forall
x,y \in A.\eqno (2.3)$$

{\bf Remark 2.2}\quad In [Se], equation (2.1) is called the operator
form of the modified classical Yang-Baxter equation on a Lie
algebra.\hfill $\Box$

In general, it is not easy to obtain ${\rm RB}(A)$ for an arbitrary
associative algebra $A$. We give some examples in certain special
cases as follows.

{\bf Example 2.3}\quad Let $A$ be a commutative associative algebra
which is the direct sum of fields. That is, there is a basis
$\{e_1,\cdots, e_n\}$ of $A$ satisfying $e_ie_j=\delta_{ij}e_j$.
Then by Rota-Baxter relation (1.1), $R=\sum\limits_{k=1}^n
r_{ik}e_k\in {\rm RB}(A)$ if and only if
$$r_{lk}r_{kl}=0,\;\;\forall\; l\ne k,$$
and
$$r_{ii}=0, r_{il}=0\;{\rm or}\;-1, l\ne i; \;\;{\rm or}\;\;
r_{ii}=1, r_{il}=0\;{\rm or}\;1, l\ne i.$$ In particular, a special
case was given in [E1] as (for any $1\leq s\leq n$)
$$R(e_i)=\sum_{l=i}^s e_l,\;1\leq i\leq s;\;\;
R(e_{s+1})=0,\;\;R(e_i)=-\sum_{l=s+1}^{i-1}e_{l},\;s+2\leq i\leq
n,$$ that is,
$$r_{ii}=1,\;r_{ij}=1,\;r_{ji}=0,\;\;1\leq i<j\leq s,
;\;\;r_{kk}=0,\;r_{kl}=-1,\;r_{lk}=0, s+1\leq l< k\leq n.$$ and
$r_{mn}=0$ in the other cases. We also list ${\rm RB}(A)$ for $n\leq
3$ in the next two sections.\hfill $\Box$

{\bf Example 2.4}\quad Let $A$ be an associative algebra in
dimension $n\geq 2$ satisfying the condition that for any two
vectors $x,y\in A$, the product $x\cdot y$ is still in the subspace
spanned by $x,y$. From [Bai], for any fixed $n\geq 2$, there are
three kinds of such (non-isomorphic) algebras. Let $\{e_1,\cdots,
e_n\}$ be a basis of $A$, then $A$ must be isomorphic to  one of the
following three algebras:

(I) $e_ie_j=0,\;\forall i,j=1,\cdots, n$;

(II) $e_1e_i=e_i, e_je_i=0,\;\;\forall i=1,\cdots, n,\;j=2,\cdots,
n$

 (III) $e_ie_1=e_i, e_ie_j=0,\;\;\forall i=1,\cdots, n,\;j=2,\cdots,
n$

\noindent It is obvious that ${\rm RB(I)}=gl(n)$ (all $n\times n$
matrices). Notice that type (III) is just type (II)' given in Lemma
2.1. Hence ${\rm RB(II)}={\rm RB(III)}$.

Moreover, we can prove that any operator $R\in {\rm RB(II)}$ if and
only if $R^2=R$. In fact, let $R(e_i)=\sum\limits_{k=1}^n r_{ik}
e_k$, then by the Rota-Baxter relation (1.1), we only need to check
the following equations (other equations hold naturally):
$$R(e_1)R(e_i)+R(e_i)=R(e_1R(e_i)+R(e_1)e_i),\;\;\forall\; \;i=1,\cdots, n.$$
For any $i$, the left hand side is $r_{11}R(e_i)+R(e_i)$ and the
right hand side is $R^2(e_i)+r_{11}R(e_i)$. Therefore, $R\in {\rm
RB(II)}$ if and only if $R^2=R$.

Furthermore, by  conclusion (5) in Lemma 2.1, we know that any
Rota-Baxter operator $R$ on the pre-Lie algebra of type (II) or type
(III) can induce an operator $N_{\alpha}=(1+\alpha)R-\alpha$
satisfying the Nijenhuis relation (2.3) for any $\alpha\in {\bf C}$.
\hfill $\Box$

On the other hand,

{\bf Definition 2.5}\quad Let $A$ be a vector space over a filed
{\bf F} with a bilinear product $(x,y)\rightarrow xy$. $A$ is called
a pre-Lie algebra if for any $x,y,z\in A$,
$$(xy)z-x(yz)=(yx)z-y(xz).\eqno (2.4)$$

It is obvious that all associative algebras are pre-Lie algebras.
For a pre-Lie algebra $A$, the commutator
$$[x,y]=xy-yx,\eqno (2.5)$$
defines a Lie algebra ${\cal G}={\cal G}(A)$, which is called the
sub-adjacent Lie algebra of $A$.

{\bf Proposition 2.6} ([GS])\quad Let $(A,\cdot)$ be an associative
algebra. If a linear operator  $R:A\rightarrow A$ satisfies the
modified Yang-Baxter equation (2.2), then the new product $*$ on $A$
given by
$$ x*y = x\cdot y + y\cdot x + [R(x), y],\;\;\forall x, y \in A \eqno (2.6)$$
defines a pre-Lie algebra.

By Proposition 2.6 and the conclusion (3) in Lemma 2.1, we can get
the following conclusion.

{\bf Corollary 2.7} Let $A$ be an associative algebra and $R$ be a
Rota-Baxter operator of weight 1. Then the product give by equation
(1.6), that is,
$$x*y=R(x)\cdot y-y\cdot R(x)-x\cdot y,\;\;\forall x,y\in A \eqno (2.7)$$
defines a pre-Lie algebra.


{\bf Definition 2.8} ([Lo]) \quad Let $A$ be a vector space over a
filed {\bf F} with two bilinear products denoted by $\prec$ and
$\succ $. $(A, \prec, \succ)$ is called a dendriform dialgebra if
for any $x,y,z\in A$,
$$(x\prec y)\prec z=x\prec (y*z),\;\;(x\succ y)\prec z=x\succ (y\prec
z),\;\;x\succ (y\succ z)=(x*y)\succ z,\eqno (2.8)$$ where
$x*y=x\prec y+x\succ y$.

{\bf Proposition 2.9} ([Ag1],[Lo])\quad Let $(A, \prec, \succ)$ be a
dendriform dialgebra. Then the product given by
$$x*y=x\prec y+x\succ y,\;\;\forall x,y\in A,\eqno (2.9)$$
defines an associative algebra ([Lo]) and the product given by
$$x\circ y=x\succ y-y\prec x,\;\;\forall x,y\in A,\eqno (2.10)$$
defines a pre-Lie algebra ([Ag1]). $(A,*)$ and $(A,\circ)$ have the
same sub-adjacent Lie algebra.

Therefore, Corollary 2.7 (and equation (1.5) and conclusion (2) in
Lemma 2.1) can also  be obtained from the following conclusion (by a
normalization of constant if necessary).

{\bf Proposition 2.10} ([Ag1], [E1])\quad Let $(A,\cdot)$ be an
associative algebra and $R$ be a Rota-Baxter operator of weight
$\lambda$, then there is a dendriform dialgebra $(A,\prec,\succ)$
defined by
$$x\prec y=x\cdot R(y)-\lambda x\cdot y,\;\;x\succ y=R(x)\cdot
y,\;\;\forall x,y\in A.\eqno (2.11)$$

It is obvious that for a commutative associative algebra $(A,\cdot)$
and any $R\in {\rm RB}(A)$, the pre-Lie algebra $(A,*)$ given by
equation (2.7) is still $(A,\cdot)$ itself. It is also obvious that
for an associative algebra $(A,\cdot)$, the pre-Lie algebra $(A,*)$
given by equation (2.7) when $R=0$ is just $(A,\cdot)$ itself and
when $R=1$ is $(A',\circ)$ given in Lemma 2.1. Moreover, we can get
a more general conclusion: Let $(A,\cdot)$ be an associative algebra
and $R\in {\rm RB}(A)$. Let $(A',\circ)$ be the associative algebra
given in Lemma 2.1. Then the pre-Lie algebra given by equation (2.7)
through $(A, R)$ is just the one given by equation (2.7) through
$(A',1-R)$.

{\bf Example 2.11}\quad Let $(A,\cdot)$ be the associative algebra
of type (II) given in Example 2.4. Then the pre-Lie algebra $(A,*)$
given by equation (2.7) satisfies
$$e_1*e_1=-e_1-\sum_{k=2}^{n}r_{1k}e_k,$$
$$e_1*e_j=(r_{11}-1)e_j,\;e_j*e_1=-\sum_{k=2}^nr_{jk}e_k, \;\;e_j*e_l=r_{j1}e_l,\forall
j,l=2,\cdots, n,$$ where $R(e_i)=\sum\limits_{k=1}^n r_{ik} e_k$ and
$R^2=R$. It is interesting that for $n=2,3$, the above pre-Lie
algebras are associative (see the next two sections). However, it is
not easy to get their classification in higher dimensions and we
have not known whether they are still associative. \hfill $\Box$

{\bf Corollary 2.12} ([EGP])\quad Let $(A,\cdot)$ be an associative
algebra and $R\in {\rm RB}(A)$, then $R$ is still a Rota-Baxter
operator of weight $\lambda=1$ on the pre-Lie algebra $(A,*)$ given
by equation (2.7).


{\bf Example 2.13}\quad Let $A$ be the 2-dimensional associative
algebra of type (II) in Example 2.4, then it is easy to see that the
operator $R$ given by $R(e_1)=e_1, R(e_2)=ae_1$ (for any $a\ne 0$)
is a Rota-Baxter operator of $A$ (also see the next section). The
pre-Lie algebra obtained by equation (2.7) is given by
$$e_1*e_1=-e_1,\;e_1*e_2=e_2*e_1=0,\;e_2*e_2=ae_2.$$
It is a commutative associative algebra  which is isomorphic to a
simple form $<e_1',e_2'|e_1'*e_1'=e_1',e_2'*e_2'=e_2'>$ (it is just
the algebra given in Example 2.3 in the case $n=2$) by a linear
transformation $e_1'\rightarrow -e_1,\;e_2'\rightarrow
\frac{1}{a}e_2$. Note that $R$ is a Rota-Baxter operator of $(A,*)$
under the same basis $\{e_1,e_2\}$ and the form  $R$ does not
satisfy the conditions given in Example 2.3. In fact, under the new
basis $\{e_1',e_2'\}$, $R$ corresponds to the new form $R'$ given by
$R'(e_1')=e_1', R'(e_2')=-e_1'$ which is consistent with the
conclusion in Example 2.3. This is an example that the matrix
presentations of Rota-Baxter operators depend on the choice of the
bases. Moreover, there is a related discussion in section 5. \hfill
$\Box$

\section
{Rota-Baxter operators on 2-dimensional associative algebras and
pre-Lie algebras}

Let $(A,\cdot)$ be an associative algebra or a pre-Lie algebra and
$\{e_1, e_2, \cdots,e_n\}$ be a basis of $A$. Let $R$ be a
Rota-Baxter operator of weight 1 on $A$. Set
$$R(e_i)=\sum _{j=1}^n r_{ij}e_{j},\;\;
e_i\cdot e_{j}=\sum_{k=i}^n C_{ij}^k e_k.\eqno (3.1)$$ Then $r_{ij}$
satisfies the following equations: $$\sum_{k,l,m=1}^n
(C_{kl}^mr_{ik}r_{jl}+C_{ij}^k
r_{km}-C_{kj}^lr_{ik}r_{lm}-C_{il}^kr_{jl}r_{km})=0,\;\;\;\forall
i,j=1,2,\cdots, n.\eqno (3.2)$$

We know that there are two 1-dimensional associative algebras ${\rm
(D0)}=<e_1|e_1e_1=0>$ and ${\rm (D1)}=<e_1|e_1e_1=e_1>$. It is easy
to see that ${\rm RB(D0)}={\bf C}$ and ${\rm
RB(D1)}=\{R|R(e_1)=0\;{\rm or}\;R(e_1)=e_1\}$.

We have known the classification of 2-dimensional complex pre-Lie
algebras ([Bu]), which includes the classification of 2-dimensional
complex associative algebras. The following results can be obtained
by  direct computation.

{\bf Proposition 3.1}\quad The Rota-Baxter operators on
2-dimensional commutative associative algebras are given in the
following table (any parameter belongs to the complex field ${\bf
C}$, unless otherwise stated).

\begin{center}
{\footnotesize \begin{tabular}{|l|l|}\hline {\normalsize Associative
algebra $A$} & {\normalsize Rota-Baxter operators RB($A$)}
\\\hline (A1) $e_1e_1=e_1,e_2e_2=e_2$ &
$\left(\begin{array}{cc} 0&0\\
0&0
\end{array}\right),
\left(\begin{array}{cc} 1&0\\
0&1
\end{array}\right),
\left(\begin{array}{cc} 0&0\\
-1&0
\end{array}\right),
\left(\begin{array}{cc} 1&0\\
1&1
\end{array}\right)$,\\
&
$\left(\begin{array}{cc} 1&0\\
0&0
\end{array}\right),
\left(\begin{array}{cc} 0&0\\
0&1
\end{array}\right),
\left(\begin{array}{cc} 0&0\\
1 & 1
\end{array}\right),
\left(\begin{array}{cc} 1&0\\
-1&0
\end{array}\right)$,\\
&$\left(\begin{array}{cc} 1&1\\
0&0
\end{array}\right),
\left(\begin{array}{cc} 0&-1\\
0&1
\end{array}\right),
\left(\begin{array}{cc} 0&-1\\
0&0
\end{array}\right),
\left(\begin{array}{cc} 1&1\\
0&1
\end{array}\right)$
\\\hline (A2) $e_2e_2=e_2,e_1e_2=e_2e_1=e_1$ &
$\left(\begin{array}{cc} 0&0\\
0&0
\end{array}\right),
\left(\begin{array}{cc} 1&0\\
0&1
\end{array}\right),
\left(\begin{array}{cc} 1&0\\
0&0
\end{array}\right),
\left(\begin{array}{cc} 0&0\\
0&1
\end{array}\right)$\\\hline
(A3) $e_1e_1=e_1$ & $\left
(\matrix{0& 0\cr 0&r_{22}\cr}\right),\left (\matrix{1& 0\cr
0&r_{22}\cr}\right)$\\\hline (A4) $e_ie_j=0$ & $\left
(\matrix{r_{11}& r_{12}\cr r_{21}&r_{22}\cr}\right)$ \\\hline (A5)
$e_1e_1=e_2$ & $\left (\matrix{r_{11}& r_{12}\cr
0&\frac{r_{11}^2}{2r_{11}-1}\cr}\right),r_{11}\ne\frac{1}{2} $
\\\hline
\end{tabular}}
\end{center}

There are two non-commutative associative algebras in dimension 2
${\rm (B1)}=<e_1|e_2e_1=e_1,e_2e_2=e_2>$ and ${\rm
(B2)}=<e_1|e_1e_2=e_1,e_2e_2=e_2>$. Both of them belong to the
algebras given in Example 2.4 in the case $n=2$, so any Rota-Baxter
operator $R$ satisfies $R^2=R$. Furthermore, we can know that (since
many of their corresponding pre-Lie algebras are isomorphic under a
basis transformation, we give a classification of these pre-Lie
algebras ``in the sense of isomorphism", that is, the corresponding
pre-Lie algebras are isomorphic to some pre-Lie algebras with
simpler presentations)

\begin{eqnarray*}
 {\rm RB(B1)}=\{ &\mbox{}& \left(\begin{array}{cc} 0&0\\
0&0
\end{array}\right)\Longrightarrow {\rm (B1)}\\
&\bigcup&\left(\begin{array}{cc} 1&0\\
0&1
\end{array}\right)\Longrightarrow {\rm (B2)}\\
&\bigcup&\left(\begin{array}{cc} 1&0\\
r_{21}&0
\end{array}\right)\Longrightarrow {\rm (A2)}\\
&\bigcup&\left(\begin{array}{cc} 0&0\\
r_{21}&1
\end{array}\right)\Longrightarrow {\rm (A3)}\\
&\bigcup&\left(\begin{array}{cc} r_{11}&r_{12}\\
r_{21}&1-r_{11}
\end{array}\right), \matrix{ r_{12} \neq 0\cr
r_{11}^2-r_{11}+r_{12}r_{21}=0\cr}\Longrightarrow {\rm (A1)}\}
\end{eqnarray*}
We also have ${\rm RB(B2)}={\rm RB(B1)}$ and the corresponding
pre-Lie algebras are given by the conclusion before Example 2.11.

{\bf Corollary 3.2}\quad Any 2-dimensional pre-Lie algebra obtained
 by equation (2.7) from a Rota-Baxter (associative) algebra is
associative.

{\bf Corollary 3.3}\quad Only the non-nilpotent commutative
associative algebras (they are (A1), (A2), (A3)) can be obtained
from 2-dimensional non-commutative associative Rota-Baxter algebras
by equation (2.7).

At the end of this section, we give the following conclusion by
direct computation.

{\bf Proposition 3.4}\quad The Rota-Baxter operators on
2-dimensional (nonassociative) pre-Lie algebras are given in the
following table.

\begin{center}
{\footnotesize \begin{tabular}{|l|l|}\hline {\normalsize Pre-Lie
algebra $A$} & {\normalsize Rota-Baxter operators RB($A$)}
\\\hline (B3)  $e_2e_1=-e_1,e_2e_2=e_1-e_2$ &
$\left(\begin{array}{cc} 0&0\\
0&0
\end{array}\right)
\left(\begin{array}{cc} 1&0\\
0&1
\end{array}\right)$
\\\hline
(B4)  $e_2e_1=-e_1,e_2e_2=ke_2, k\neq -1$ & $k\neq 0: \matrix{\left(\begin{array}{cc} 0&0\\
0&0
\end{array}\right)
\left(\begin{array}{cc} 1&0\\
0&1
\end{array}\right)\cr
\left(\begin{array}{cc} 0&0\\
0&1
\end{array}\right)
\left(\begin{array}{cc} 0&0\\
0&1
\end{array}\right)\cr
}$\\
\cline{2-2}
&$k= 0: \matrix{
\left(\begin{array}{cc} 1&0\\
0&r_{22}
\end{array}\right)
\left(\begin{array}{cc} 0&0\\
r_{21}&0
\end{array}\right)r_{21}\neq 0
\cr
\left(\begin{array}{cc} 0&0\\
0&r_{22}
\end{array}\right)
\left(\begin{array}{cc} 1&0\\
r_{21}&1
\end{array}\right)r_{21}\neq 0\cr
}$
\\\hline
(B5)  $e_1e_2=le_1,e_2e_1=(l-1)e_1,e_2e_2=e_1+le_2,l\neq 0$ &
$l=1: \left(\begin{array}{cc} 0&0\\
0&0
\end{array}\right)
\left(\begin{array}{cc} 1&0\\
0&1
\end{array}\right)$\\
\cline{2-2}
&$l\neq 1:\matrix{ \left(\begin{array}{cc} 0&0\\
0&0
\end{array}\right)
 \left(\begin{array}{cc} 1&0\\
\frac{1}{l-1}&0
\end{array}\right)
\cr
\left(\begin{array}{cc} 1&0\\
0&1
\end{array}\right)
\left(\begin{array}{cc} 0&0\\
-\frac{1}{l-1}&1
\end{array}\right)
\cr }$
\\\hline
 (B6)  $e_1e_1=2e_1,e_1e_2=e_2,e_2e_2=e_1$ &
$\left(\begin{array}{cc} 0&0\\
0&0
\end{array}\right)
\left(\begin{array}{cc} 1&0\\
0&1
\end{array}\right)$
\\\hline
\end{tabular}}
\end{center}

Since (B6) is the unique simple pre-Lie algebra (without any ideals
besides zero and itself) in dimension 2 ([Bu]), we have

{\bf Corollary 3.5}\quad There is no non-trivial Rota-Baxter
operator on the 2-dimensional simple pre-Lie algebra, that is, only
$0,1$ are the Rota-Baxter operators.

\section{Rota-Baxter operators on 3-dimensional associative algebras
and their corresponding pre-Lie algebras}

It is easy to get the classification of 3-dimensional complex
associative algebras (for example, see [LHB]). Then by  direct
computation, we have the following results.

{\bf Proposition 4.1}\quad The Rota-Baxter operators on
3-dimensional commutative associative algebras are given in the
following table.

\begin{center}
{\footnotesize \begin{tabular}{|l|l|}\hline {\normalsize Associative
algebra $A$} & {\normalsize Rota-Baxter operators RB($A$)}
\\\hline
(C1) $e_ie_j=0$ & $ \left(\matrix{r_{11}& r_{12}&r_{13}\cr
r_{21}&r_{22}&r_{23}\cr r_{31}&r_{32}&r_{33}\cr}\right)$
\\\hline
(C2) $e_3e_3=e_1$ &  $\left(\begin{array}{ccc} r_{11}&0&0\\
r_{21}&r_{22}&0\\
r_{31}&r_{32}&r_{11}\pm\sqrt{r_{11}^2-r_{11}}
\end{array}\right)$\\\hline
 (C3) $\left\{\matrix{ e_2e_2=e_1\cr
e_3e_3=e_1\cr}\right.$ &
$\left(\begin{array}{ccc} r_{11}&0&0\\
r_{21}&r_{11}&0\\
r_{31}&0&r_{11}
\end{array}\right)r_{11}=0, 1$ \\&
$\left(\begin{array}{ccc} r_{11}&0&0\\
r_{21}&r_{22}&r_{23}\\
r_{31}&-r_{23}&r_{22}
\end{array}\right), \matrix{r_{23}\neq
0,\cr r_{22}=r_{11}\pm\sqrt{r_{11}^2-r_{11}-r_{23}^2}\cr} $\\&
$\left(\begin{array}{ccc} r_{11}&0&0\\
r_{21}&r_{22}^-&r_{23}\\
r_{31}&r_{23}&r_{22}^+
\end{array}\right)\matrix{r_{23}\neq
0,\cr r_{22}^+ =r_{11}+\sqrt{r_{11}^2-r_{11}-r_{23}^2}\cr r_{22}^-
=r_{11}-\sqrt{r_{11}^2-r_{11}-r_{23}^2}\cr}$ \\&
$\left(\begin{array}{ccc} r_{11}&0&0\\
r_{21}&r_{22}^+&r_{23}\\
r_{31}&r_{23}&r_{22}^-
\end{array}\right)\matrix{r_{23}\neq
0,\cr r_{22}^+ =r_{11}+\sqrt{r_{11}^2-r_{11}-r_{23}^2}\cr r_{22}^-
=r_{11}-\sqrt{r_{11}^2-r_{11}-r_{23}^2}\cr}$
\\\hline
 (C4)
$\left\{\matrix{e_2e_3=e_3e_2=e_1\cr e_3e_3=e_2\cr}\right.$ &
$\left(\begin{array}{ccc}\frac{r_{22}r_{33}}{r_{22}+r_{33}-1}&0&0\\
r_{21}&r_{22}&0\\
r_{31}&r_{32}&r_{33}
\end{array}\right),\matrix{r_{33}=r_{22}\pm\sqrt{r_{22}^2-r_{22}}\cr
r_{21}=\frac{2r_{32}r_{33}(1-r_{33})}{(1-2r_{33})(r_{22}+r_{33}-1)}\cr}
$
\\\hline (C5)
$\left\{\matrix{e_1e_1=e_1\cr e_2e_2=e_2\cr e_3e_3=e_3\cr}\right.$&
$\left(\begin{array}{ccc} r_{11}&0&0\\
0&r_{22}&0\\
0&0&r_{33}
\end{array}\right)$
$\left(\begin{array}{ccc} r_{11}&0&0\\
0&r_{22}&0\\
2r_{33}-1&2r_{33}-1&r_{33}
\end{array}\right)$ \\&
$\left(\begin{array}{ccc} r_{11}&0&0\\
0&r_{22}&0\\
2r_{33}-1&0&r_{33}
\end{array}\right)$
$\left(\begin{array}{ccc} r_{11}&0&0\\
0&r_{22}&0\\
0&2r_{33}-1&r_{33}
\end{array}\right)$ \\ &

$\left(\begin{array}{ccc} r_{11}&0&0\\
2r_{22}-1&r_{22}&0\\
0&0&r_{33}
\end{array}\right)$
$\left(\begin{array}{ccc} r_{11}&0&0\\
2r_{22}-1&r_{22}&0\\
2r_{33}-1&2r_{33}-1&r_{33}
\end{array}\right)$ \\ &

$\left(\begin{array}{ccc} r_{11}&2r_{11}-1&0\\
0&r_{22}&0\\
0&0&r_{33}
\end{array}\right)$
$\left(\begin{array}{ccc} r_{11}&2r_{11}-1&0\\
0&r_{22}&0\\
2r_{33}-1&2r_{33}-1&r_{33}
\end{array}\right)$ \\&

$\left(\begin{array}{ccc} r_{11}&2r_{11}-1&2r_{11}-1\\
0&r_{22}&0\\
0&0&r_{33}
\end{array}\right)\ $
$\left(\begin{array}{ccc} r_{11}&2r_{11}-1&2r_{11}-1\\
0&r_{22}&0\\
0&2r_{33}-1&r_{33}
\end{array}\right)\ $\\&

$\left(\begin{array}{ccc} r_{11}&0&0\\
2r_{22}-1&r_{22}&2r_{22}-1\\
0&0&r_{33}
\end{array}\right)\ $
$\left(\begin{array}{ccc} r_{11}&0&0\\
2r_{22}-1&r_{22}&2r_{22}-1\\
2r_{33}-1&0&r_{33}
\end{array}\right)\ $\\&

$\left(\begin{array}{ccc} r_{11}&0&0\\
0&r_{22}&2r_{22}-1\\
0&0&r_{33}
\end{array}\right)\ $

$\left(\begin{array}{ccc} r_{11}&2r_{11}-1&2r_{11}-1\\
0&r_{22}&2r_{22}-1\\
0&0&r_{33}
\end{array}\right)\ $\\&

$\left(\begin{array}{ccc} r_{11}&0&2r_{11}-1\\
0&r_{22}&0\\
0&0&r_{33}
\end{array}\right)$
$\left(\begin{array}{ccc} r_{11}&0&2r_{11}-1\\
2r_{22}-1&r_{22}&2r_{22}-1\\
0&0&r_{33}
\end{array}\right)$ \\ & $r_{11}=0, 1, r_{22}=0, 1, r_{33}=0, 1$\\ \hline

(C6) $\left\{\matrix{ e_2e_2=e_2\cr e_3e_3=e_3}\right.$ &$\left(\begin{array}{ccc} r_{11}&0&0\\
0&r_{22}&0\\
0&r_{32}&0
\end{array}\right), r_{22}=0, 1 , r_{32}=0, -1$\\
&$\left(\begin{array}{ccc} r_{11}&0&0\\
0&r_{22}&0\\
0&r_{32}&1
\end{array}\right), r_{22}=0, 1 , r_{32}=0, 1$\\&
$\left(\begin{array}{ccc} r_{11}&0&0\\
0&r_{22}&2r_{22}-1\\
0&0&r_{33}
\end{array}\right),r_{22}=0, 1 , r_{33}=0, 1$
\\\hline
\end{tabular}}
\end{center}

\begin{center}
{\footnotesize \begin{tabular}{|l|l|}\hline {\normalsize Associative
algebra $A$} & {\normalsize Rota-Baxter operators
RB($A$)\hspace{3.8cm}}
\\\hline
(C7) $\left\{\matrix{ e_1e_3=e_3e_1=e_1 \cr e_2e_2=e_2\cr
e_3e_3=e_3\cr}\right.$ &
$\left(\begin{array}{ccc} r_{11}&0&0\\
0&r_{22}&0\\
0&r_{32}&0
\end{array}\right),r_{11}=0, 1 ,r_{22}=0, 1 , r_{32}=0, -1$\\&
$\left(\begin{array}{ccc} r_{11}&0&0\\
0&r_{22}&0\\
0&r_{32}&1
\end{array}\right),r_{11}=0, 1 ,r_{22}=0, 1 , r_{32}=0, 1$\\
&$\left(\begin{array}{ccc} r_{11}&0&0\\
0&0&-1\\
0&0&r_{33}
\end{array}\right)r_{11}=0, 1 ,r_{33}=0, 1 $\\
&$\left(\begin{array}{ccc} r_{11}&0&0\\
0&1&1\\
0&0&r_{33}
\end{array}\right),r_{11}=0, 1 ,r_{33}=0, 1$
\\\hline
(C8) $e_3e_3=e_3$ & $\left(\begin{array}{ccc} r_{11}&r_{12}&0\\
r_{21}&r_{22}&0\\
0&0&r_{33}
\end{array}\right), r_{33}=0, 1$\\\hline
(C9) $\left\{\matrix{ e_1e_3=e_3e_1=e_1
\cr e_3e_3=e_3\cr}\right.$ &$\left(\begin{array}{ccc} r_{11}&r_{12}&0\\
0&r_{22}&0\\
0&0&1-r_{11}
\end{array}\right)r_{12}\neq 0, r_{11}=0,1$\\
&$\left(\begin{array}{ccc} r_{11}&0&0\\
0&r_{22}&0\\
0&0&r_{33}
\end{array}\right), r_{11}=0, 1, r_{33}=0, 1$\\
&$\left(\begin{array}{ccc} r_{11}&0&0\\
r_{21}&r_{22}&0\\
0&0&r_{11}
\end{array}\right),r_{11}=0, 1, r_{21}\neq 0$
\\\hline
(C10) $\left\{\matrix{ e_1e_3=e_3e_1=e_1 \cr e_2e_3=e_3e_2=e_2\cr
e_3e_3=e_3\cr}\right.$ &$\left(\begin{array}{ccc} r_{11}&r_{12}&0\\
r_{21}&1-r_{11}&0\\
0&0&r_{33}
\end{array}\right),\matrix{r_{12}\neq 0 \cr  r_{33}=0, 1\cr
r_{11}-r_{11}^2-r_{12}r_{21}=0\cr}$\\
&$\left(\begin{array}{ccc} r_{11}&0&0\\
0&r_{11}&0\\
0&0&r_{33}
\end{array}\right),r_{33}=0, 1, r_{11}=0, 1$\\
& $
\left(\begin{array}{ccc} 1&0&0\\
r_{21}&0&0\\
0&0&r_{33}
\end{array}\right), r_{33}=0, 1$\\
&$\left(\begin{array}{ccc} 0&0&0\\
r_{21}&1&0\\
0&0&r_{33}
\end{array}\right),r_{33}=0, 1$\\\hline
(C11) $\left\{\matrix{ e_1e_1=e_2 \cr e_3e_3=e_3\cr}\right.$ &
$\left(\begin{array}{ccc} r_{22}\pm\sqrt{r_{22}^2-r_{22}}&r_{12}&0\\
0&r_{22}&0\\
0&0&r_{33}
\end{array}\right),r_{33}=0, 1$
\\\hline
(C12) $\left\{\matrix{ e_1e_1=e_2\cr e_1e_3=e_3e_1=e_1 \cr
e_2e_3=e_3e_2=e_2\cr e_3e_3=e_3\cr}\right.$ &$\left(\begin{array}{ccc} r_{11}&0&0\\
0&r_{11}&0\\
0&0&r_{33}
\end{array}\right),r_{33}=0, 1, r_{11}=0, 1$\\\hline
\end{tabular}}
\end{center}


{\bf Proposition 4.2}\quad The Rota-Baxter operators on
3-dimensional non-commutative associative algebras and their
corresponding pre-Lie algebras given by equation (2.7) (in the sense
of isomorphism) are given in the following table.

\begin{center}
{\footnotesize \begin{tabular}{|l|l|l|}\hline {\normalsize
Associative algebra $A$} & {\normalsize Rota-Baxter operators
RB($A$)} & {\normalsize Pre-Lie algebra }
\\\hline
 (T1)$\left\{ \matrix {e_1 \cdot e_2 = \frac{1}{2} e_3\cr
e_2 \cdot e_1 = -\frac{1}{2} e_3\cr} \right.$&$
\left(\begin{array}{ccc} r_{11}&r_{12}&r_{13}\\
r_{21}&1-r_{11}&r_{23}\\
0&0&r_{33}
\end{array}\right)$, $r_{11}^2-r_{11}+r_{12}r_{21}=0$&(T1)(C3)\\
\cline{2-3} &$
\left(\begin{array}{ccc} r_{11}&r_{12}&r_{13}\\
r_{21}&r_{22}&r_{23}\\
0&0&\frac{r_{11}r_{22}-r_{12}r_{21}}{r_{11}+r_{22}-1}
\end{array}\right)$, $r_{11}+r_{22}-1 \neq 0$&$\matrix{{\rm (T1), (T2),}\cr {\rm (T3)}_\lambda, \lambda\ne 0\cr}$\\
\cline{2-3} \hline
\end{tabular}}
\end{center}

\begin{center}
{\footnotesize \begin{tabular}{|l|l|l|}\hline {\normalsize
Associative algebra $A$} & {\normalsize Rota-Baxter operators
RB($A$)} & {\normalsize Pre-Lie algebra }
\\\hline
(T2) $e_2 \cdot e_1 = - e_3$&$
\left(\begin{array}{ccc} 0&0&r_{13}\\
0&1&r_{23}\\
0&0&r_{33}
\end{array}\right)$&(C1)\\
\cline{2-3} &$
\left(\begin{array}{ccc} 1&0&r_{13}\\
0&0&r_{23}\\
0&0&r_{33}
\end{array}\right)$&(T1)\\
\cline{2-3} &$
\left(\begin{array}{ccc} r_{11}&0&r_{13}\\
0&r_{22}&r_{23}\\
0&0&\frac{r_{11}r_{22}}{r_{11}+r_{22}-1}
\end{array}\right)r_{11}+r_{22}-1 \neq 0$&(T2), (T3)$_\lambda, \lambda\ne 0$\\
\cline{2-3} &$
\left(\begin{array}{ccc} 0&0&r_{13}\\
r_{21}&1&r_{23}\\
0&0&1
\end{array}\right)(r_{21}\neq 0)$, $
\left(\begin{array}{ccc} 0&r_{12}&r_{13}\\
0&1&r_{23}\\
0&0&0
\end{array}\right)(r_{12}\neq 0)$
&(C2)\\
\cline{2-3} &$
\left(\begin{array}{ccc} 1&0&r_{13}\\
r_{21}&0&r_{23}\\
0&0&0
\end{array}\right)(r_{21}\neq 0)$, $
\left(\begin{array}{ccc} 1&r_{12}&r_{13}\\
0&0&r_{23}\\
0&0&1
\end{array}\right)(r_{12}\neq 0)$&(C3)\\
\cline{2-3} &$
\left(\begin{array}{ccc} 0&r_{12}&r_{13}\\
0&r_{22}&r_{23}\\
0&0&0
\end{array}\right)r_{12}\neq 0,r_{22}\neq 1$& (T2)\\
\cline{2-3} &$
\left(\begin{array}{ccc} 1&r_{12}&r_{13}\\
0&r_{22}&r_{23}\\
0&0&1
\end{array}\right)r_{12}\neq 0,r_{22}\neq 0$&(T2), (T3)$_\lambda, \lambda\ne 0$\\
\cline{2-3} &$
\left(\begin{array}{ccc} r_{11}&0&r_{13}\\
r_{21}&0&r_{23}\\
0&0&0
\end{array}\right)r_{11}\neq 1,r_{21}\neq 0$&(T2), (T3)$_\lambda, \lambda\ne 0$\\
\cline{2-3} &$
\left(\begin{array}{ccc} r_{11}&0&r_{13}\\
r_{21}&1&r_{23}\\
0&0&1
\end{array}\right)r_{11}\neq 0,r_{21}\neq 0$&(T2)\\
\cline{2-3} \hline (T3)$_\lambda$$\matrix{\left\{ \matrix{e_1 \cdot
e_1 =  e_3\cr e_1 \cdot e_2 = e_3\cr e_2 \cdot e_2 =\lambda e_3\cr}
\right.\cr \lambda \neq 0\cr}$
&$ \left(\begin{array}{ccc} r_{11}&r_{12}&r_{13}\\
r_{21}&r_{22}&r_{23}\\
r_{31}&r_{32}&r_{33}
\end{array}\right)$ &(T3)$_\lambda$, $\lambda \neq 0$\\&
{\small$\matrix{ r_{11}^2+r_{11}(r_{12}-2r_{33})+(\lambda
r_{12}^2+r_{33}-r_{12}r_{33})=0;\cr
r_{11}r_{21}+r_{11}r_{22}+\lambda r_{12}r_{22}+r_{33}(1-r_{11}\cr
-\lambda r_{12}-r_{21}-r_{22})=0;\cr
r_{21}r_{11}+r_{12}r_{21}+\lambda r_{22}r_{12}-r_{21}r_{33}-\lambda
r_{12}r_{33}=0;\cr r_{21}^2+r_{21}(r_{22}-r_{33})+(\lambda
r_{22}^2+\lambda r_{33}-2\lambda r_{22}r_{33})=0\cr}$}&
\\\hline
 (T4)$\left\{
\matrix{ e_3 \cdot e_2 = e_2\cr e_3 \cdot e_3 =e_3\cr} \right.$&$
\left(\begin{array}{ccc} r_{11}&r_{12}&0\\
0&1&0\\
0&0&1
\end{array}\right)r_{12}\neq 0$&(T5)\\
\cline{2-3} &$
\left(\begin{array}{ccc} r_{11}&r_{12}&0\\
0&0&0\\
0&0&0
\end{array}\right)r_{12}\neq 0$&(N1)$\left\{\matrix{ e_1*e_3=e_2\cr e_3*e_1=e_2\cr
e_3*e_2=e_2\cr e_3*e_3=e_3\cr}\right.$ \\
\cline{2-3} &$
\left(\begin{array}{ccc} r_{11}&0&0\\
0&0&0\\
0&0&0
\end{array}\right)$&(T4)\\
\cline{2-3} &$
\left(\begin{array}{ccc} r_{11}&0&0\\
r_{21}&1&0\\
r_{21}r_{32}&r_{32}&0
\end{array}\right)$&(C9)\\
\cline{2-3} &$
\left(\begin{array}{ccc} r_{11}&0&0\\
r_{21}&0&0\\
-r_{21}r_{32}&r_{32}&1
\end{array}\right)$&(C8)\\
\cline{2-3} &$
\left(\begin{array}{ccc} r_{11}&0&0\\
0&1&0\\
0&0&1
\end{array}\right)$&(T5)\\
\cline{2-3}  \hline
\end{tabular}}
\end{center}

\begin{center}
{\footnotesize \begin{tabular}{|l|l|l|}\hline {\normalsize
Associative algebra $A$} & {\normalsize Rota-Baxter operators
RB($A$)} & {\normalsize Pre-Lie algebra }\\\hline &$
\left(\begin{array}{ccc} r_{11}&0&0\\
0&r_{22}&r_{23}\\
0&r_{32}&1-r_{22}
\end{array}\right)$, $\matrix{ r_{23}\neq0\cr
r_{22}^2-r_{22}+r_{23}r_{32}=0\cr}$&(C6)\\
\cline{2-3}\hline (T5) $\left\{ \matrix{e_2 \cdot e_3 =e_2\cr e_3
\cdot e_3 =e_3\cr} \right. $& RB(T4)&
The same as in (T4)\\
(=(T4)')&&through Corollary 2.12\\\hline (T6)$\left\{ \matrix{ e_1
\cdot e_1 =e_1 \cr e_3 \cdot e_2 =e_2 \cr e_3 \cdot e_3 =e_3} \right.$&$\left(\begin{array}{ccc} r_{11}&0&0\\
0&0&0\\
r_{31}&0&0
\end{array}\right),r_{11}=0,1, r_{31}=0, -1$&(T6)\\
\cline{2-3} &$
\left(\begin{array}{ccc} r_{11}&0&0\\
0&1&0\\
r_{31}&0&1
\end{array}\right),r_{11}=0, 1, r_{31}=0, 1$&(T7)\\
\cline{2-3} &$
\left(\begin{array}{ccc} r_{11}&0&0\\
0&1&0\\
r_{31}&r_{32}&0
\end{array}\right),r_{11}=0,1, r_{31}=0,-1$&(C7)\\
\cline{2-3} &$
\left(\begin{array}{ccc} r_{11}&0&0\\
0&0&0\\
r_{31}&r_{32}&1
\end{array}\right),r_{11}=0, 1, r_{31}=0, 1$&(C6)\\
\cline{2-3} &$
\left(\begin{array}{ccc} r_{11}&0&0\\
0&r_{22}&r_{23}\\
0&r_{32}&1-r_{22}
\end{array}\right),\matrix{r_{23}\neq 0\cr r_{11}=0, 1\cr
r_{22}^2-r_{22}+r_{23}r_{32}=0\cr}$&(C5)\\
\cline{2-3} &$
\left(\begin{array}{ccc} r_{11}&0&0\\
r_{23}&r_{22}&r_{23}\\
-r_{22}&r_{32}&1-r_{22}
\end{array}\right), \matrix{r_{23}\neq 0\cr r_{11}=0, 1\cr
r_{22}^2-r_{22}+r_{23}r_{32}=0\cr}$&(C5)\\
\cline{2-3} &$
\left(\begin{array}{ccc} r_{11}&0&0\\
r_{23}&r_{22}&r_{23}\\
1-r_{22}&r_{32}&1-r_{22}
\end{array}\right),\matrix{r_{23}\neq 0\cr r_{11}=0, 1\cr
r_{22}^2-r_{22}+r_{23}r_{32}=0\cr}$&(C5)\\
\cline{2-3} &$
\left(\begin{array}{ccc} 0&r_{12}&-1\\
0&0&0\\
0&0&0
\end{array}\right)$&(N2)$\left\{ \matrix{ e_1*e_1=e_1+2e_3\cr
e_1*e_3=-e_3\cr e_3*e_1=-e_3\cr e_3*e_2=e_2\cr e_3e_3=e_3\cr}\right.$\\
\cline{2-3} &$
\left(\begin{array}{ccc} 1&r_{12}&1\\
0&0&0\\
0&0&0
\end{array}\right)$&(N3)$\left\{ \matrix{ e_1*e_1=e_1\cr
e_1*e_3=e_3\cr e_3*e_1=e_3\cr e_3*e_2=e_2\cr e_3e_3=e_3\cr}\right.$
\\\cline{2-3}
&$
\left(\begin{array}{ccc} 0&r_{12}&-1\\
0&1&0\\
0&0&1
\end{array}\right)$&(T9)\\
\cline{2-3} &$
\left(\begin{array}{ccc} 1&r_{12}&1\\
0&1&0\\
0&0&1
\end{array}\right)$&(N4) $\left\{ \matrix{ e_1*e_1=e_1\cr
e_1*e_2=e_2\cr e_2*e_1=e_2\cr
e_1*e_3=e_3\cr e_3*e_1=e_3\cr e_3*e_2=e_2\cr e_3e_3=-e_3\cr}\right.$\\
\cline{2-3} &$
\left(\begin{array}{ccc} 0&r_{12}&-1\\
0&1&0\\
0&r_{12}&0
\end{array}\right)$&(N5) $\left\{ \matrix{ e_1*e_1=e_1\cr
e_1*e_2=e_2\cr e_2*e_1=e_2\cr
e_3*e_2=e_2\cr e_3e_3=e_3\cr}\right.$\\
\cline{2-3} &$
\left(\begin{array}{ccc} 1&r_{12}&1\\
0&1&0\\
0&-r_{12}&0
\end{array}\right)$&(N6)$\left\{ \matrix{ e_1*e_1=e_1\cr
e_1*e_2=e_2\cr e_2*e_1=e_2\cr
e_3*e_2=e_2\cr e_3e_3=-e_3\cr}\right.$\\
\cline{2-3}  \hline
\end{tabular}}
\end{center}

\begin{center}
{\footnotesize \begin{tabular}{|l|l|l|}\hline {\normalsize
Associative algebra $A$} & {\normalsize Rota-Baxter operators
RB($A$)} & {\normalsize Pre-Lie algebra }\\\hline &$
\left(\begin{array}{ccc} 0&r_{12}&-1\\
0&0&0\\
0&-r_{12}&1
\end{array}\right)$&(T5)\\
\cline{2-3} &$
\left(\begin{array}{ccc} 1&r_{12}&1\\
0&0&0\\
0&r_{12}&1
\end{array}\right)$&(N7) $\left\{ \matrix{ e_1*e_1=e_1\cr
e_3*e_2=e_2\cr e_3e_3=-e_3\cr}\right.$\\
\hline (T7)$\left\{ \matrix{ e_1 \cdot e_1 =e_1 \cr e_2 \cdot e_3
=e_2 \cr e_3 \cdot e_3 =e_3} \right.$& RB(T6)&
The same as in (T6)\\
(=(T6)')&&through Corollary 2.12\\\hline (T8)$\left\{ \matrix{e_1
\cdot e_3 =e_1\cr e_3 \cdot e_1 =e_1\cr e_3 \cdot e_2 =e_2\cr e_3
\cdot e_3 =e_3\cr} \right.$&$
\left(\begin{array}{ccc} r_{11}&0&0\\
0&0&0\\
0&0&0
\end{array}\right), r_{11}=0,1$&(T8)\\
\cline{2-3} &$
\left(\begin{array}{ccc} r_{11}&0&0\\
0&1&0\\
0&0&1
\end{array}\right), r_{11}=0, 1$&(N8)$\left\{ \matrix{ e_1*e_3=e_1\cr e_3*e_1=e_1\cr
e_2*e_3=e_2\cr e_3e_3=e_3\cr}\right.$\\
\cline{2-3} &$
\left(\begin{array}{ccc} r_{11}&0&0\\
0&1&0\\
0&r_{32}&0
\end{array}\right), r_{11}=0,1$&(C10)\\
\cline{2-3} &$
\left(\begin{array}{ccc} r_{11}&0&0\\
0&0&0\\
0&r_{32}&1
\end{array}\right),r_{11}=0, 1$&(C9)\\
\cline{2-3} &$
\left(\begin{array}{ccc} r_{11}&0&0\\
0&r_{22}&r_{23}\\
0&r_{32}&1-r_{22}
\end{array}\right),\matrix{r_{23}\neq 0\cr r_{11}=0, 1\cr r_{22}^2-r_{22}+r_{23}r_{32}=0\cr}$&(C7)\\
\cline{2-3} &$
\left(\begin{array}{ccc} 0&0&0\\
r_{21}&1&0\\
r_{21}r_{32}&r_{32}&0
\end{array}\right), r_{21}\neq 0$&(C10)\\
\cline{2-3} &$
\left(\begin{array}{ccc} 1&0&0\\
r_{21}&0&0\\
-r_{21}r_{32}&r_{32}&1
\end{array}\right), r_{21}\neq 0$&(C9)\\
\cline{2-3} &$
\left(\begin{array}{ccc} 0&r_{12}&0\\
0&1&0\\
0&0&1
\end{array}\right)r_{12}\neq 0$&(N9)$\left\{ \matrix{e_1*e_3=-e_1+e_2\cr
e_2*e_3=-e_2\cr e_3*e_1=-e_1+e_2\cr e_3*e_3=-e_3\cr}\right.$
\\
\cline{2-3} &$
\left(\begin{array}{ccc} 1&r_{12}&0\\
0&0&0\\
0&0&0
\end{array}\right), r_{12}\neq 0$&(T8)\\
\cline{2-3} \hline (T9)$\left\{ \matrix{ e_1 \cdot e_1 =e_1\cr e_1
\cdot e_2 =e_2\cr e_1 \cdot e_3 =e_3\cr e_2 \cdot e_1 =e_2\cr
 e_3 \cdot e_1 =e_3\cr e_3 \cdot e_2 =e_2\cr  e_3 \cdot
e_3 =e_3} \right.$&$
\left(\begin{array}{ccc} 0&0&0\\
0&0&0\\
0&0&0
\end{array}\right)$,$
\left(\begin{array}{ccc} 1&0&0\\
0&1&0\\
0&0&1
\end{array}\right)$&(T9)\\
\cline{2-3} &$
\left(\begin{array}{ccc} 0&0&0\\
0&1&0\\
0&r_{32}&0
\end{array}\right)$&(C7)\\ \cline{2-3} &$
\left(\begin{array}{ccc} 1&0&0\\
0&0&0\\
r_{31}&0&0
\end{array}\right), r_{31}=0, 1$&(T9)\\
\cline{2-3}&$
\left(\begin{array}{ccc} 1&0&0\\
0&1&0\\
r_{31}&r_{32}&0
\end{array}\right),r_{31}=0, 1$ &(C7)\\
\cline{2-3}\hline
\end{tabular}}
\end{center}

\begin{center}
{\footnotesize \begin{tabular}{|l|l|l|}\hline {\normalsize
Associative algebra $A$} & {\normalsize Rota-Baxter operators
RB($A$)} & {\normalsize Pre-Lie algebra }\\\hline  &$
\left(\begin{array}{ccc} 0&0&0\\
0&0&0\\
r_{31}&r_{32}&1
\end{array}\right), r_{31}=0,-1$&(C7)\\
\cline{2-3} &$
\left(\begin{array}{ccc} 1&0&0\\
0&0&0\\
0&r_{32}&1
\end{array}\right)$&(C7)\\
\cline{2-3} &$
\left(\begin{array}{ccc} 0&0&0\\
0&1&0\\
r_{31}&0&1
\end{array}\right), r_{31}=0,-1$&(T9)\\\cline{2-3} &$
\left(\begin{array}{ccc} r_{11}&0&0\\
0&r_{22}&r_{23}\\
0&r_{32}&1-r_{22}
\end{array}\right),\matrix{ r_{23}\neq 0\cr r_{11}=0, 1\cr
r_{22}^2-r_{22}+r_{23}r_{32}=0\cr}$&(C5)\\
\cline{2-3} &$
\left(\begin{array}{ccc} 0&0&0\\
-r_{23}&r_{22}&r_{23}\\
r_{22}-1&r_{32}&1-r_{22}
\end{array}\right),\matrix{ r_{23}\neq 0\cr r_{22}^2-r_{22}+r_{23}r_{32}=0\cr}$&(C5)\\
\cline{2-3} &$
\left(\begin{array}{ccc} 1&0&0\\
-r_{23}&r_{22}&r_{23}\\
r_{22}&r_{32}&1-r_{22}
\end{array}\right),\matrix{ r_{23}\neq 0\cr r_{22}^2-r_{22}+r_{23}r_{32}=0\cr}$&(C5)\\
\cline{2-3} &
$\left(\begin{array}{ccc} r_{11}&0&-1\\
0&r_{22}&0\\
0&0&0
\end{array}\right), r_{22}=0, 1, r_{11}=0, 1$&(N5)\\
\cline{2-3} &$
\left(\begin{array}{ccc} r_{11}&0&1\\
0&r_{22}&0\\
0&0&1
\end{array}\right), r_{22}=0, 1, r_{11}=0,1$&(T7)\\
\cline{2-3} &$
\left(\begin{array}{ccc} 2&0&-1\\
0&r_{22}&0\\
1&0&0
\end{array}\right),r_{22}=0, 1$&(N5)\\
\cline{2-3} &$
\left(\begin{array}{ccc} -1&0&1\\
0&r_{22}&0\\
-1&0&1
\end{array}\right), r_{22}=0,1$&(T7)\\
\cline{2-3}&
$\left(\begin{array}{ccc} r_{11}&r_{12}&-1\\
0&0&0\\
0&0&0
\end{array}\right),r_{11}=0,1, r_{12}\neq 0$&(N5)\\
\cline{2-3} &$
\left(\begin{array}{ccc} r_{11}&r_{12}&-1\\
0&1&0\\
0&r_{12}&0
\end{array}\right), r_{11}=0,1, r_{12}\neq 0$&(N5)\\
\cline{2-3} &$
\left(\begin{array}{ccc} r_{11}&r_{12}&1\\
0&0&0\\
0&r_{12}&1
\end{array}\right), r_{11}=0, 1, r_{12}\neq 0$&(T7)\\
\cline{2-3} &$
\left(\begin{array}{ccc} r_{11}&r_{12}&1\\
0&1&0\\
0&0&1
\end{array}\right), r_{11}=0, 1, r_{12}\neq 0$& (T7)\\
\cline{2-3} &$
\left(\begin{array}{ccc} 2&r_{12}&-1\\
0&0&0\\
1&0&0
\end{array}\right), r_{12}\neq 0$&(N5) \\
\cline{2-3} &$
\left(\begin{array}{ccc} 2&r_{12}&-1\\
0&1&0\\
1&r_{12}&0
\end{array}\right), r_{12}\neq 0$&(N5) \\
\cline{2-3} &$
\left(\begin{array}{ccc} -1&r_{12}&1\\
0&0&0\\
-1&r_{12}&1
\end{array}\right), r_{12}\neq 0$&(T7)\\
\cline{2-3} &$
\left(\begin{array}{ccc} -1&r_{12}&1\\
0&1&0\\
-1&0&1
\end{array}\right), r_{12}\neq 0$&(T7)\\
\cline{2-3}\hline
\end{tabular}}
\end{center}

\begin{center} {\footnotesize \begin{tabular}{|l|l|l|}\hline
$\matrix{{\rm  Associative}\cr {\rm algebra}\; A\cr}$ & {\normalsize
Rota-Baxter operators RB($A$)} & {\normalsize Pre-Lie algebra
}\\\hline (T10)$\left\{ \matrix{ e_3 \cdot e_1 =e_1\cr e_3 \cdot e_2
=e_2\cr e_3 \cdot e_3 =e_3\cr}\right.$ & $\{ R|R^2=R\}$&
$\matrix{{\rm (T6),(T8),(T9),}\cr {\rm (T10),(T11),(C10)}\cr}$\\
\hline (T11)$\left\{ \matrix{ e_1 \cdot e_3 =e_1\cr e_2 \cdot e_3
=e_2\cr e_3 \cdot e_3 =e_3\cr}\right.$ & $\{ R|R^2=R\}$&
$\matrix{{\rm (T6),(T8),(T9),}\cr {\rm (T10),(T11),(C10)}\cr}$\\
\hline (T12)$\left\{ \matrix{e_3 \cdot e_1 =e_1\cr e_2 \cdot e_3
=e_2\cr e_3 \cdot e_3 =e_3\cr} \right.$&$
\left(\begin{array}{ccc} 0&0&0\\
0&0&0\\
0&0&0
\end{array}\right)$, $
\left(\begin{array}{ccc} 1&0&0\\
0&1&0\\
0&0&1
\end{array}\right)$&(T12)\\
\cline{2-3}  &$
\left(\begin{array}{ccc} 1&0&0\\
0&0&0\\
r_{31}&0&0
\end{array}\right)$&(N8)\\
\cline{2-3} &$
\left(\begin{array}{ccc} 1&0&0\\
0&1&0\\
r_{31}&r_{32}&0
\end{array}\right)$&(C9)\\
\cline{2-3} &$
\left(\begin{array}{ccc} 1&0&0\\
0&0&0\\
0&r_{32}&1
\end{array}\right)$& (N8)\\
\cline{2-3} &$
\left(\begin{array}{ccc} 0&0&0\\
0&1&0\\
0&r_{32}&0
\end{array}\right)$&(T4)\\
\cline{2-3} &$
\left(\begin{array}{ccc} 0&0&0\\
0&0&0\\
r_{31}&r_{32}&1
\end{array}\right)$&(C9)\\
\cline{2-3} &$
\left(\begin{array}{ccc} 0&0&0\\
0&1&0\\
r_{31}&0&1
\end{array}\right)$&(T4)\\
\cline{2-3} &$
\left(\begin{array}{ccc} 0&0&0\\
r_{21}&1&0\\
r_{21}r_{32}&r_{32}&0
\end{array}\right), r_{21}\neq 0$&(N1) \\
\cline{2-3} &$
\left(\begin{array}{ccc} 1&0&0\\
r_{21}&0&0\\
-r_{21}r_{32}&r_{32}&1
\end{array}\right), r_{21}\neq 0$&(N9)
\\
\cline{2-3}   &$
\left(\begin{array}{ccc} r_{11}&0&0\\
0&r_{22}&r_{23}\\
0&r_{32}&1-r_{22}
\end{array}\right)$ $\matrix{ r_{23}\neq 0\cr r_{11}=0, 1\cr r_{22}^2-r_{22}+r_{23}r_{32}=0\cr}$
& $\matrix{ {\rm (N10)} \left\{ \matrix{ e_1*e_1=e_1\cr
e_1*e_3=e_3\cr e_3*e_1=e_3\cr e_3*e_2=e_2\cr
e_3*e_3=-e_3\cr}\right.\cr
{\rm }\cr}$\\ && (N6) \\
\cline{2-3}  &$
\left(\begin{array}{ccc} r_{11}&0&0\\
r_{21}&r_{22}&r_{23}\\
\frac{r_{21}(1-r_{11}-r_{22})}{r_{23}}&r_{32}&1-r_{22}
\end{array}\right), \matrix{ r_{23}\neq 0\cr r_{21}\neq 0\cr r_{11}=0, 1\cr
r_{22}^2-r_{22}+r_{23}r_{32}=0\cr}$ &  (N6), (N10) \\
\cline{2-3} &$
\left(\begin{array}{ccc} r_{11}&0&r_{13}\\
0&r_{22}&0\\
r_{31}&0&1-r_{11}
\end{array}\right),\matrix{ r_{13}\neq 0\cr r_{22}=0, 1\cr r_{11}^2-r_{11}+r_{13}r_{31}=0\cr}$&(N6), (N10)\\
\cline{2-3} &$
\left(\begin{array}{ccc} r_{11}&r_{12}&r_{13}\\
0&r_{22}&0\\
r_{31}&\frac{r_{12}(1-r_{11}-r_{22})}{r_{13}}&1-r_{11}
\end{array}\right),\matrix{ r_{13}\neq 0\cr r_{12}\neq 0\cr r_{22}=0, 1\cr
r_{11}^2-r_{11}+r_{13}r_{31}=0\cr}$ &(N6), (N10)\\
\cline{2-3} &$
\left(\begin{array}{ccc} 1&r_{12}&0\\
0&0&0\\
r_{31}&r_{31}r_{12}&0
\end{array}\right), r_{12}\neq 0$&(N9)\\
\cline{2-3} &$
\left(\begin{array}{ccc} 0&r_{12}&0\\
0&1&0\\
r_{31}&-r_{31}r_{12}&1
\end{array}\right), r_{12}\neq 0$&(N1)\\
\cline{2-3}\hline
\end{tabular}}
\end{center}


{\bf Corollary 4.4}\quad The algebras of type (N1)-(N10) are the
only nonassociative pre-Lie algebras obtained from 3-dimensional
Rota-Baxter algebras.

{\bf Corollary 4.5}\quad The sub-adjacent Lie algebras of the
nonassociative pre-Lie algebras obtained from 3-dimensional
Rota-Baxter algebras are unique up to isomorphism:
$$<e_1,e_2,e_3 |[ e_2, e_3 ]= e_2>.$$
 It is the direct sum of the 2-dimensional
non-abelian Lie algebra and 1-dimensional center.

{\bf Corollary 4.6}\quad Besides the algebras of type (C4), (C11)
and (C12), the 3-dimensional commutative associative algebras can be
obtained from noncommutative associative Rota-Baxter algebras by
equation (2.7).

\section{Discussion and conclusions}

From the study in the previous sections, we give the following
discussion and conclusions.

(1) We have given all the Rota-Baxter operators of weight 1 on
complex associative algebras in dimension $\leq 3$. They can help us
to construct pre-Lie algebras. We would like to point out that such
constructions have some constraints. For example, all the pre-Lie
algebras obtained from 2-dimensional Rota-Baxter algebras are
associative and the sub-adjacent Lie algebras of the nonassociative
pre-Lie algebras obtained from 3-dimensional Rota-Baxter algebras
are unique up to isomorphism.

(2) By conclusion (3) in Lemma 2.1, the Rota-Baxter operators that
we obtained in this paper can help us to get the examples of
operators satisfying (the operator form of) the modified classical
Yang-Baxter equation in the sub-adjacent Lie algebras of these
associative algebras.

(3) It is hard and less practicable to extend our study to be in
higher dimensions since the Rota-Baxter relation involves the
nonlinear quadratic equations (3.2). Moreover, for a Rota-Baxter
algebra $A$, both the set ${\rm RB}(A)$ and the corresponding
pre-Lie algebras obtained from $A$ rely on the choice of a basis of
$A$ and its corresponding structural constants (see Example 2.13).
So it might be enough to search some interesting examples (not
necessary to get the whole set ${\rm RB}(A)$) in higher dimensions,
even in infinite dimension ([E1]). In this sense, our study can be a
good guide (like Examples 2.3-2.4).

(4) The construction in Corollary 2.7 cannot be extended to the
nonassociative pre-Lie algebras, that is, we cannot obtain pre-Lie
algebras from a nonassociative Rota-Baxter pre-Lie algebra by
equation (2.7). However, if the induced pre-Lie algebra
$(A,*)=(A,*_1)$ from a Rota-Baxter (associative) algebra $(A,\cdot,
R)$ is still associative, then $(A,*_1, R)$ is still a Rota-Baxter
algebra which can induce a new pre-Lie algebra $(A,*_2)$ with $R$
being a Rota-Baxter operator (it is also a double construction, see
[EGK] and [LHB]). Therefore, we can get a series of Rota-Baxter
(associative) algebras $(A,*_n, R)$ for any $n\in {\bf N}$ or there
exists some $N\in {\bf N}$ such that $(A,*_n,R)$ is a Rota-Baxter
associative algebra for any $n<N$ and $(A,*_N,R)$ is a
nonassociative Rota-Baxter pre-Lie algebra.

(5) We  have also given the Rota-Baxter operators of weight 1 on
2-dimensional complex pre-Lie algebras. It is interesting to
consider certain geometric structures related to these examples and
the possible application in physics.

\section*{Acknowledgements}

The authors thank Professor Li Guo and the referees' valuable
suggestion. In particular, the authors are grateful of being told
the close relations between dendriform dialgebras and pre-Lie
algebras.  This work was supported in part by the National Natural
Science Foundation of China (10571091, 10621101), NKBRPC
(2006CB805905)£¬ and Program for New Century Excellent Talents in
University.


\baselineskip=14pt
\begin{thebibliography}{}
\item[[AS]] A. Andrada, S. Salamon, Complex product structures on Lie
algebras,  Forum Math. 17 (2005) 261-295.
\item[[Ag1]] M. Aguiar, Pre-Poisson algebras, Lett. Math. Phys. 54
(2000) 263-277.
\item[[Ag2]] M. Aguiar, Infinitesimal Hopf algebras, Contemporary
Mathematics 267, Amer. Math. Soc., (2000) 1-29.
\item[[Ag3]] M. Aguiar, On the associative analog of Lie bialgebras,
J. Algebra 244 (2001), no. 2, 492-532.
\item[[At]] F.V. Atkinson, Some aspects of Baxter's functional
equation, J. Math. Anal. Appl. 7 (1963) 1-30.
\item[[Au]] Auslander, L. Simply transitive groups of affine motions.
Amer. J. Math. 99 (1977), no. 4,  809-826.
\item[[Bai]] C.M. Bai,
Left-symmetric algebras from linear functions, J. Algebra 281
(2004), no. 2,  651-665.
\item[[BM1]] C.M. Bai, D.J. Meng, On the realization of transitive Novikov
algebras, J. Phys. A: Math. Gen. 34 (2001), no. 16,  3363-3372.
\item[[BM2]] C.M. Bai, D.J. Meng, The realizations of non-transitive Novikov
algebras, J. Phys. A: Math. Gen. 34 (2001), no.33,  6435-6442.
\item[[BM3]] C.M. Bai, D.J. Meng, A Lie algebraic approach to Novikov
algebras, J. Geom. Phys. 45 (2003), no. 1-2,  218-230.
\item[[BK]] B. Bakalov, V. Kac, Field algebras, Int. Math. Res. Not.
(2003), no. 3, 123-159.
\item[[BN]] A.A. Balinskii, S.P. Novikov, Poisson brackets of hydrodynamic
type, Frobenius algebras and Lie algebras, Soviet Math. Dokl. 32
(1985) 228-231.
\item[[Bax]] G. Baxter, An analytic problem whose solution follows
from a simple algebraic identity, Pacific J. Math. 10 (1960)
731-742.
\item[[Bo]] M. Bordemann, Generalized Lax pairs,
the modified classical Yang-Baxter equation, and affine geometry of
Lie groups, Comm. Math. Phys. 135 (1990), no. 1, 201-216.
\item[[Bu]] D. Burde, Simple left-symmetric algebras with solvable Lie
algebra, Manuscipta Math. 95 (1998), no. 3, 397-411.
\item[[CGM]]J. Carinena, J. Grabowski, G. Marmo, Quantum
bi-Hamiltonian systems, Int. J. Modern Phys. A 15 (2000), no. 30,
4797-4810.
\item[[Ca]] P. Cartier, On the structure of free Baxter
algebras, Advances in Math. 9 (1972) 253-265.
\item[[CL]] F. Chapoton, M. Livernet, Pre-Lie algebras and the rooted trees
operad, Int. Math. Res. Not. (2001), no. 8,  395-408.
\item[[Ch]] B.Y. Chu, Symplectic homogeneous spaces, Trans. Amer. Math. Soc.
197 (1974) 145-159.
\item[[CK1]] A. Connes, D. Kreimer, Hopf
 algebras, renormalization and noncommutative geometry,
 Comm. Math. Phys. 199 (1998), no. 1, 203-242.
\item[[CK2]] A. Connes, D. Kreimer, Renormalization in quantum field
theory and the Riemann-Hilbert problem. I. The Hopf algebra
structure of graphs and the main theorem,  Comm. Math. Phys. 210
(2000), no. 1, 249-273.
\item[[CK3]] A. Connes, D. Kreimer, Renormalization in quantum field
theory and the Riemann-Hilbert problem. II. The $\beta$-function,
diffeomorphisms and the renormalization group, Comm. Math. Phys. 216
(2001), no. 1, 215-241.
\item[[CK4]] A. Connes, D. Kreimer, Insertion
and Elimination: the doubly infinite Lie algebra of Feynman graphs,
Annales Henri Poincare, 3 (2002) no. 3, 411-433.
\item[[Deb]] S.Luiz, de Braganca, Finite dimensional Baxter algebras,
Studies in Applied Math. 54 (1975), no. 1, 75-89.
\item[[Der]] N.A. Derzko, Mappings satisfying Baxter's identity in
the algebra of matrices, J. Math. Anal. Appl. 42 (1973) 1-19.
\item[[E1]] K. Ebrahimi-Fard, Loday-type algebras and the
Rota-Baxter relation, Lett. Math. Phys. 61 (2002), no. 2, 139-147.
\item[[E2]] K. Ebrahimi-Fard, On the associative Nijenhuis relation,
Elect. J. Comb.,11 (2004), no. 1, Research Paper 38.
\item[[EGP]] K. Ebrahimi-Fard, J.M. Gracia-Bondia, F. Patras,
Rota-Baxter algebras and new combinatorial identities,
arXiv:math.CO/0701031.
\item[[EG1]] K. Ebrahimi-Fard,  L. Guo, Rota-Baxter algebras and dendriform algebras, arXiv:math/0503647.
\item[[EG2]] K. Ebrahimi-Fard,  L. Guo, On free Rota-Baxter algebras, arXiv:math.RA/0510266.
\bibitem[EGK]{Fard2} K. Ebrahimi-Fard,  L. Guo, D. Kreimer, Integrable renormalization 1: The ladder case,
 J. Math. Phys, 45 (2004) , No. 10, 3758-3769.
\item[[ES]] P. Etingof, A. Soloviev, Quantization of geometric classical
$r$-matrix, Math. Res. Lett. 6 (1999), no. 2, 223-228.
\item[[FG]] H. Figueroa, J.M. Gracia-Bondia, Combinatorial Hopf
algebras in qunantum field theory I, Rev. Math. Phys. 17 (2005), no.
8,  881-976.
\item[[GD]] I.M. Gel'fand,  I. Ja. Dorfman, Hamiltonian operators and
algebraic structures associated with them, Funktsional. Anal. i
Prilozhen. 13 (1979), no. 4, 13-30, 96.
\item[[G]] M. Gerstenhaber, The cohomology structure of an
associative ring, Ann. of Math (2). 78 (1963) 267-288.
\item[[GS]] I.Z. Golubchik, V.V. Sokolov, Generalized operator Yang-Baxter
equations, integrable ODEs and nonassociative algebras, J. Nonlinear
Math. Phys., 7 (2000), no.2, 184-197.
\item[[GK1]] L. Guo, W. Keigher, Baxter algebras and shuffle
products, Adv. Math. 150 (2000), no. 1, 117-149.
\item[[GK2]] L. Guo, W. Keigher, On free Baxter algebras:
completions and the internal construction, Adv. Math. 151 (2000),
no. 1, 101-127.
\item[[Ki]] H. Kim, Complete left-invariant affine structures on nilpotent
Lie groups, J. Differential Geometry 24 (1986), no. 3, 373-394.
\item[[Ku1]] B.A. Kupershmidt, Non-abelian phase spaces, J. Phys. A: Math.
Gen. 27 (1994), no. 8, 2801-2809.
\item[[Ku2]] B.A. Kupershmidt, On the nature of the Virasoro algebra, J.
Nonlinear Math. Phy., 6 (1999), no. 2, 222-245.
\item[[Ku3]] B.A. Kupershmidt, What a classical $r$-matrix really is, J.
Nonlinear Math. Phy., 6 (1999), no. 4, 448-488.
\item[[Le1]] P. Leroux, Construction of Nijenhuis operators and
dendriform trialgebras, arXiv:math.QA/0503647.
\item[[Le2]] P. Leroux, Ennea-algebras, J. Algebra 281 (2004), no.
1, 287-302.
\item[[LHB]] X.X. Li, D.P. Hou, C.M. Bai, Rota-Baxter operators on pre-Lie
algebras, J. Nonlinear Math. Phy., 14 (2007), no. 2, 269-289.
\item[[Lo]] J.-L. Loday, Dialgebras, in Dialgebras and related
operads, Lecture Notes in Math. 1763 (2002) 7-66.
\item[[LR]] J.-L. Loday, M. Ronco, Trialgebras and families of
polytopes, in ``Homotopy Theory: Relations with Algebraic Geometry,
Group Cohomology, and Algebraic K-theory'', Comtep. Math. 346 (2004)
369-398.
\item[[Me]] A. Medina, Flat left-invariant connections adapted to the
automorphism structure of a Lie group, J. Differential Geometry 16
(1981), no. 3, 445-474.
\item[[Mi1]] J.B. Miller, Some properties of Baxter operators, Acta
Math. Acad. Sci. Hungar. 17 (1966) 387-400.
\item[[Mi2]] J.B. Miller, Baxter operators and
endomorphisms on Banach algebras, J. Math. Anal. Appl. 25 (1969)
503-520.
\item[[N]] Nguyen-Huu-Bong, Some apparent connection between Baxter
and averaging operators, J. Math. Anal. Appl. 56 (1976), no. 2,
330-345.
\item[[R1]] G.-C. Rota, Baxter algebras and combinatorial
identities I, Bull. Amer. Math. Soc. 75 (1969) 325-329.
\item[[R2]] G.-C. Rota, Baxter algebras and combinatorial
identities II, Bull. Amer. Math. Soc. 75 (1969) 330-334.
\item[[R3]] G.-C. Rota, Baxter operators, an introduction,
In: ``Gian-Carlo Rota on Combinatorics, Introductory papers and
commentaries", Joseph P.S. Kung, Editor, Birkh\"auser, Boston, 1995.
\item[[R4]] G.-C. Rota, Ten mathematics problems I will never
solve, Mitt. Dtsch. Math,-Ver.  (1998), no. 2, 45-52.
\item[[Se]] M.A. Semenov-Tyan-Shanskii, What is a classical R-matrix?
Funct. Anal. Appl. 17 (1983) 259-272.
\item[[Sh]] H. Shima, Homogeneous Hessian manifolds, Ann. Inst. Fourier (Grenoble) 30 (1980), no. 3,  91-128.
\item[[Sp]] F. Spitzer, A combinatorial lemma and its application to
probability theory, Trans. Amer. Math. Soc. 82 (1956) 323-339.
\item[[SS]] S.I. Svinolupov, V.V. Sokolov, Vector-matrix generalizations of
classical integrable equations, Theoret. and Math. Phys. 100 (1994),
no. 2, 959-962.
\item[[V]] E.B. Vinberg, The theory of   homogeneous convex cones, Transl. of Moscow Math.
Soc. No. 12 (1963) 303-358.
\item[[Z]] E.I. Zel'manov,
On a class of local translation invariant Lie algebras, Soviet Math.
Dokl. 35 (1987) 216-218.
\end{thebibliography}
\end{document}